\documentclass[acmsigecom]{acmtrans2m}

\usepackage{amsmath,amssymb,amsfonts,xspace,xcolor,multirow}

\makeatletter\let\bibhang\@undefined\makeatother
\usepackage{natbib}
\usepackage{algorithm}
\usepackage{algorithmic}
\usepackage{booktabs}
\def\newblock{\hskip .11em plus .33em minus .07em}
\usepackage{paralist}

\newcommand{\Pref}[1][]{
	\ifthenelse{\equal{#1}{}}{\mathrel R}{\mathop{R_{#1}}}
}                                          
\newcommand{\sPref}[1][]{                  
	\ifthenelse{\equal{#1}{}}{\mathrel P}{\mathop{P_{#1}}}
}                                          
\newcommand{\Indiff}[1][]{                 
	\ifthenelse{\equal{#1}{}}{\mathrel I}{\mathop{I_{#1}}}
}
\newcommand{\prefset}[1][]{\ifthenelse{\equal{#1}{}}{\mathcal{R}}{\mathcal{R}_{#1}}}

\newdef{definition}{Definition}

\title{Justifications of Welfare Guarantees under Normalized Utilities}
\author{HARIS AZIZ}
\begin{abstract}
It is standard in computational social choice to analyse welfare considerations under the assumptions of normalized utilities. In this note, we summarize some common reasons for this approach. We then mention another justification which is ignored but has solid normative appeal. The central concept used in the `new' justification can also be used more widely as a social objective.  
\end{abstract}
            
            
\begin{document}

\begin{bottomstuff}
Authors' addresses: \texttt{haris.aziz@unsw.edu.au}
\end{bottomstuff}

\maketitle
            
\section{Introduction}

% Boutilier, C., Caragiannis, I., Haber, S., Lu, T., Procaccia, A.D., Sheffet, O.: Optimal social choice functions: a utilitarian view. In: Proceedings of the 13th ACM Conference on Electronic Commerce, pp. 197–214. ACM (2012)

% Filos-Ratsikas, A., Miltersen, P.B.: Truthful approximations to range voting. CoRR, abs/1307.1766 (2013)

Social welfare under normalized utilities is frequently considered in the computational social choice literature. It is especially the case in multi-agent resource allocation, mechanism design without money and utilitarian voting. The welfare notions considered include utilitarian social welfare (sum of agents' utilities), egalitarian social welfare (minimum of agents' utilities) and Nash social welfare  (product of agents' utilities). Most of the focus is on utilitarian welfare which goes back to ideas of \citet{Bent89}. 

%All three stands are thriving research topics in the field. 

 % and mechanism design without money. 

For multi-agent resource allocation under ordinal preferences, researchers have considered how well social welfare is approximated by specific mechanisms for truthful preferences or preferences in equilibrium~(see e.g., \citep{RKFZ14a,GuCo10a,BFT11a}). The standard assumption in these papers is that utilities are normalized i.e., an agent's sum of utilities for all items is one. The assumption has been termed as \emph{unit-sum}.%\footnote{Another version of normalized utilities is \emph{unit-range} whereby the maximum utility is one and minimum utility for an item is zero.}

Similarly, there is growing literature on \emph{implicit utilitarian voting}~(see e.g., \citep{CNPS16a,LuBo11b,BCH+12a}) where normalized utilities are popular. For these settings, the \emph{distortion} of a voting rule is used as a measure of how good the rule is. Distortion
is the worst ratio of the maximum  utilitarian social welfare versus the utilitarian social welfare achieved among all problem instances. Many results concerning distortion bounds  hinge on the assumption the utilities are normalized.\footnote{There is another stream of results on distortion that assume that utilities are induced by a metric~(see e.g., \citep{AnPo16a}.)}

 %Again, in many of the papers, it is typically assumed that the utilities are normalized. 

In other multi-agent resource allocation problems, agents are asked to express cardinal utilities over items but these utilities are assumed to be normalized or are processed to be normalized. For example, \citet{BoLe15a} state that
most collective utility functions only make sense if the utilities are expressed on a common scale or normalized. 
Even when agent preferences are ordinal, the scoring rules used to get `proxy utilities' satisfy the normalized or equal scale property~(see e.g., \citep{BBL14a}). 
%In several works, the goal is to maximize social welfare where social welfare could be utilitarian, egalitarian, or Nash~\citep{NNRJ14a}. 

In short, both within voting and resource allocation, there is a focus on welfare under normalized utilities.\footnote{The welfare notion of utilitarian social welfare under normalized utilities has been referred to as `relative utilitarianism'~\citep{DhMe99a}.} When the `real' utilities of agents are not normalized, it begs the question that why normalized utilities are used. The papers making this assumption either justify it as a standard assumption used in previous work or by mentioning one or two reasons. In this note, we curate and discuss justifications for welfare approaches under normalized utilities.  %The following are more standard reasons for considering normalized utilities. 

\paragraph*{\textbf{First Principle/Philosophical Justifications}}

\begin{enumerate}
	\item \textbf{Scale invariance}. Egalitarian and utilitarian social welfare suffer from being responsive to scale variance.\footnote{Nash welfare on the other hand is scale invariant. The Spliddit website~\citep{GoPr14a} which provides a convenient interface for fair division algorithms uses a solution based on the Nash welfare~\citep{CKM+16a} for allocation of goods.}
	 %Even Nash social welfare suffers from the issue when outcomes are discrete. 
Imposing  normalized utilities is a simple way to regain scale invariance. 
\item \textbf{Guaranteeing a proportion of the ultimate happiness of an individual}. Another way to view scale invariance is that one is concerned less about the amount of utility achieved by an individual agent and more so about the fraction of maximum possible utility that she achieves. This concern is indeed captured when one tries to approximate egalitarian welfare under normalized utilities.
	\item \textbf{Fairness}. In voting, when each voter is allowed to spread a utility of 1 over the alternatives, it is akin to each voter having `one' vote and `equal' say. Similarly, normalized utilities are considered in resource allocation to define rules such as Adjusted Winner~\citep{BrTa96a}. %,ABFF15a
If utilities are not normalized, and the rule is welfarist in some sense, then fairness can be undermined. For example, when considering rules that are responsive to utilitarian social welfare, the outcome is favourable to agents with the most magnified utilities. When considering egalitarian welfare, the allocation is aligned with the concerns of the agent with most scaled down utility valuations. 
	\end{enumerate}
	
	\paragraph*{\textbf{Justification in a common setting}}
	
	\begin{enumerate}
	\item[(4)] \textbf{Probabilistic perspective.} Another possible justification for using normalized utilities is using the utility as a measure of certainty of liking an item or outcome. 
	\end{enumerate}
	
	\paragraph*{\textbf{Technical Justifications}}
	\begin{enumerate}
	\item[(5)] \textbf{Strategic}. The reasons for fairness can also be seen as justification regarding strategic issues. Considering normalized utilities circumvents certain  trivial manipulation actions of scaling up or down of utilities.  
When considering utilitarian welfare normalization prevents agents from overshadowing other agents by reporting extremely high utilities. An agent who reports the highest utilities for all items would get all the items in a utilitarian social welfare maximizing solution!
	% \item Fairness justification. Again the above reason can be used in terms of justification for fairness.

	\item[(6)] \textbf{Reasonable approximation guarantees for welfare}. Another reason for considering normalized utilities  is also technical and somewhat `self-serving'. There is little hope of achieving reasonable approximation guarantees of maximum welfare or reasonable low distortion when utilities are unbounded. Therefore the normalization assumption can be seen as a trick of the trade to obtain reasonable approximation guarantees.  

	\end{enumerate}

We have discussed reasons for considering (approximate) welfare guarantee for normalized utilities.
There are at least two possible points of criticism concerning social welfare under normalized utilities. Firstly, one may negatively perceive that the welfare guarantee results only hold under the restrictive assumption of normalizes utilities. Secondly,
there can be a more classical objection to a cardinal approach to social welfare itself which involves interpersonal comparison of utilities~(see e.g., \citep{Robb35}). The second issue may appear especially acute in the context of social choice settings such as fair allocation and voting which typically do not involve money. One reason for pursuing social welfare in computer science research  has been that `without explicit optimization objective that measures the quality of outcomes, approximation cannot play a role'~\citep{PrTe13a}. %There is a third criticism as well which applies to models where agents are required report precise cardinal utilities for outcomes which can be cognitively challenging or even impossible. 

We point out that by simply pursuing justification (6), one can paradoxically get a  justification that avoids both criticisms. 
One can get a similar guarantee for a concept that does not involve interpersonal comparison of utilities and does not require scaling of utilities of any agent. 
This largely ignored `new' justification is  centered around a relaxation of Pareto optimality.

% \medskip
\begin{enumerate}
	\item[(7)] \textbf{Approximate Pareto optimality under any scaling of the utilities}. 
\end{enumerate}

We say that utility profile $u=(u_1,\ldots, u_n)$ Pareto dominates utility profile $u'=(u_1',\ldots, u_n')$ if $u_i\geq u_i'$ for all agents $i$ and $u_i> u_i'$ for some agent $i$.
Given any $\alpha\in [0,1]$, a utility profile $u$ is \emph{$\alpha$-Pareto optimal} if there exists no other achievable utility profile $u'$ such that $\alpha\cdot u'$ Pareto dominates $u$.\footnote{The definition is written for positive utilities but can be adapted for negative or mixed utilities.} The concept is very natural and has been used in the context of routing games~\citep{AuDo10a}, probabilistic matchings~\citep{ILWM17a} and participatory budgeting~\citep{ABM17a}.

It can be proven that any outcome that achieves $\alpha$ fraction of the maximum social welfare under normalized utilities also satisfies $\alpha$-Pareto optimality under any scaling of the utilities. 
Suppose agents have normalized utilities. Consider any outcome that achieves $\alpha$ fraction of the maximum social welfare. Suppose the utility profile of the agents is $u$. Then we first claim that $u$ is $\alpha$-Pareto optimal. Suppose it is not $\alpha$-Pareto optimal. Then there exists another achievable utility profile $u'$ such that $\alpha\cdot u'$ Pareto dominates $u$. But this means that $u$ achieves less than $\alpha$ fraction of social welfare of $u'$ 
which is a contradiction.
We have established that the outcome achieves $\alpha$-Pareto optimality under normalized utilities. We note that $\alpha$-Pareto optimality is invariant under scaling of an agent's utility function even when discrete outcomes are considered. 
The reason is that for an agent $i$, any two utility functions $u_i$ and $u_i'$ where $u_i'=\alpha{u_i}$ and for any social outcomes $a,b$, $u_i(a)\geq u_i(b)$ if and only if $u_i'(a)\geq u_i'(b)$. Hence, the outcome achieves $\alpha$-Pareto optimality under any scaling of the agents' utilities.

\medskip
By normalizing the utilities and achieving or establishing some bound on the maximum utilitarian or egalitarian social welfare also implies the same approximation bound on Pareto optimality for the `real' utilities of the agents. Thus one can use justification (6) for technical ease but then achieve a guarantee that has more wide-spread normative appeal.  The `new' justification can also be used as another motivation for the projects of implicit utilitarian voting and approximate mechanism design without money. %~\citep{PrTe13a}
It can also be viewed as a
source of corollaries for these lines of work. 
Taking another view, for new settings, one can directly use approximate Pareto optimality rather than particular social welfare as the social objective. A lower bound result for approximate Pareto optimality would imply a similar lower bound for utilitarian welfare. 

\medskip
We conclude by mentioning that the normative/axiomatic approach in traditional social choice
and the quantitive welfarist approaches employed in recent computational social choice papers have been termed as distinct from each other. 
The use of approximate Pareto optimality provides a convenient bridge between the
two approaches.

\section*{Acknowledgments}

The author thanks Felix Brandt, Hu Fu, Xin Huang, Barton Lee, Ariel Procaccia and Matt Weinberg for useful comments.

   % \small \bibliography{../../pamas/abbshort,../../pamas/haris_master,../../pamas/brandt,../../pamas/aziz}

\end{document}